\documentclass{emulateapj}

\usepackage{ifthen}
\newboolean{color}

\setboolean{color}{false}
\setboolean{color}{true}

\def \th{\thinspace}

\def \eg{{{\it e.g.},\ }}
\def \etal{{\it et al.\ }}

\def \ie{{{\it i.e.}\ }}

\def \vs{{\it vs.\ }}
\def \Teff{{$T_{\rm {ef\!f}} $}}
\def \teff{{T_{\rm {ef\!f}} }}

\def\Mo{{$M_\odot $}}
\def\ML{{$M$-$L$\ }}
\def \aanda{A\&A}

\long\def\jumpover#1{{}}

\slugcomment{Submitted to ApJ}

\begin{document}

\title{Beat Cepheids as Probes of Stellar and Galactic Metallicity}
\author{J.~Robert Buchler\altaffilmark{1}
and
R\'obert Szab\'o\altaffilmark{1}}
\altaffiltext{1}{Physics Department, University of Florida,
Gainesville, FL 32611, USA}

\begin{abstract} 

The mere location of a Beat Cepheid model in a Period Ratio \vs Period diagram
(Petersen diagram) puts very tight constraints on its metallicity $Z$.  The
Beat Cepheid Peterson diagrams are revisited with linear nonadiabatic turbulent
convective models, and their accuracy as a probe for stellar metallicity is
evaluated.  They are shown to be largely independent of the helium content $Y$,
and they are also only weakly dependent on the mass-luminosity relation that is
used in their construction.  However, they are found to show sensitivity to the
relative abundances of the elements that are lumped into the metallicity
parameter $Z$.  Rotation is estimated to have but a small effect on the
'pulsation metallicities'.  A composite Petersen diagram is presented that
allows one to read off upper and lower limits on the metallicity $Z$ from the
measured period $P_0$ and period ratio $P_{10}$.

\end{abstract}





\maketitle

\section{Introduction} \label{sec:intro}

The idea of using the periods extracted from the lightcurves of Beat Cepheids
({\bf BC}s) to constrain the astrophysical parameters of these Cepheids is
certainly not new.  It goes back to the introduction of Period Ratio \vs Period
diagrams, or \cite{petersen73} diagrams ({\bf PD}s). The usefulness of PDs
comes from the fact that the period ratios $P_{10}=P_1/P_0$ turn out to be
strongly dependent on the metallicity $Z$.  This was subsequently exploited to
show that the newly introduced OPAL opacities \citep{iglesias96} largely
removed a longstanding mass discrepancy problem (\eg \cite{mbm}).  With the
microlensing projects a large number of extra-galactic BCs were discovered, and
the consequent utility of PDs, both observationally, \eg
\citep{andrievsky,bbk01}, and theoretically, \eg
\citep{morganwelch,cdp95,kovacs} was quickly realized.

The modeling of galactic evolution relies heavily on observational galactic
metallicity distributions. The latter have traditionally been estimated by
spectroscopic means of a number of different objects: through the observations
of HII regions (\eg \cite{vilchez88}, \cite{urbaneja05}), of B type supergiant
stars \citep{monteverde97} of Planetary Nebulae \citep{magrini04}, of Wolf
Rayet stars \citep{abbott04}, of red giant branch photometry \citep{tiede04}
and of classical Cepheids \citep{andrievsky}.  Techniques based on multiband
photometry of regular Cepheid lightcurves \citep{caputo01} and the use of the
shapes of Bump Cepheid lightcurves \citep{kw06} have also been proposed.  All these
methods have varying degrees of accuracy and, in the case of M33, for example,
a controversy has arisen between the 'traditional' large value of the [O/H]
gradient $\sim -0.11$ dex/kpc \citep{vilchez88,garnett97} and a revised
shallower one of --0.012\th dex/kpc of \citet{crockett06}.

It is therefore extremely useful to have an additional, independent and
accurate method for obtaining a metallicity.  In this spirit Beaulieu, Buchler,
Marquette, Hartman \& Schwarzenberg (2006) used 5 newly discovered BCs in M33
as metallicity tracers.  From the 'pulsational' metallicities that they
calculated and from the galacto-centric distances of these Cepheids they could
extract a metallicity gradient of --0.16\th dex/kpc for M33 which allowed them
to weigh heavily in favor of the traditional value of [O/H] and against a
recent downward revision.

In this paper we explore in some detail both the power and the limitations of
the 'pulsational' determination of the metallicities of BCs on the basis of
linear nonadiabatic convective stellar models.  In a subsequent paper we will
present an extended survey of nonlinear models, and we show how they can be
used to further narrow down the metallicity determination of BCs.

\begin{figure*}
\epsscale{1}
\ifthenelse{\boolean{color}}
{\plotone{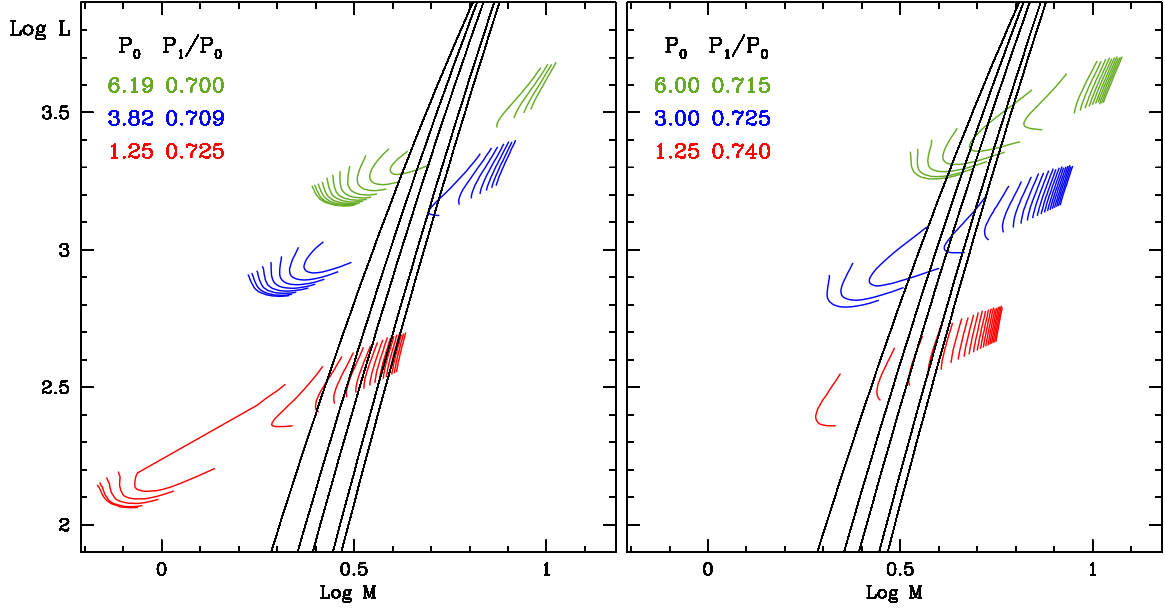}}
{\plotone{plotloci_bw.ps}}
\caption{Loci of potential BCs as a function of $Z$ for specified pairs.
Left: ($P_0, P_{10}$) =  (6.188, 0.7005),(3.827, 0.7091), (1.250, 0.7250);  
right: = (6.00, 0.7150),  (3.00, 0.7250), (1.250, 0.7400).  $Z$
increases by 0.001 between the successive loci. The vertical lines represent
\ML relations that are derived from \cite{piet} for $Z$=0.001, 0.004, 0.008,
0.019, 0.027, from left to right.}
\label{plotloci}
\end{figure*}

\begin{figure*}
\epsscale{1.0}
\ifthenelse{\boolean{color}}
{\plotone{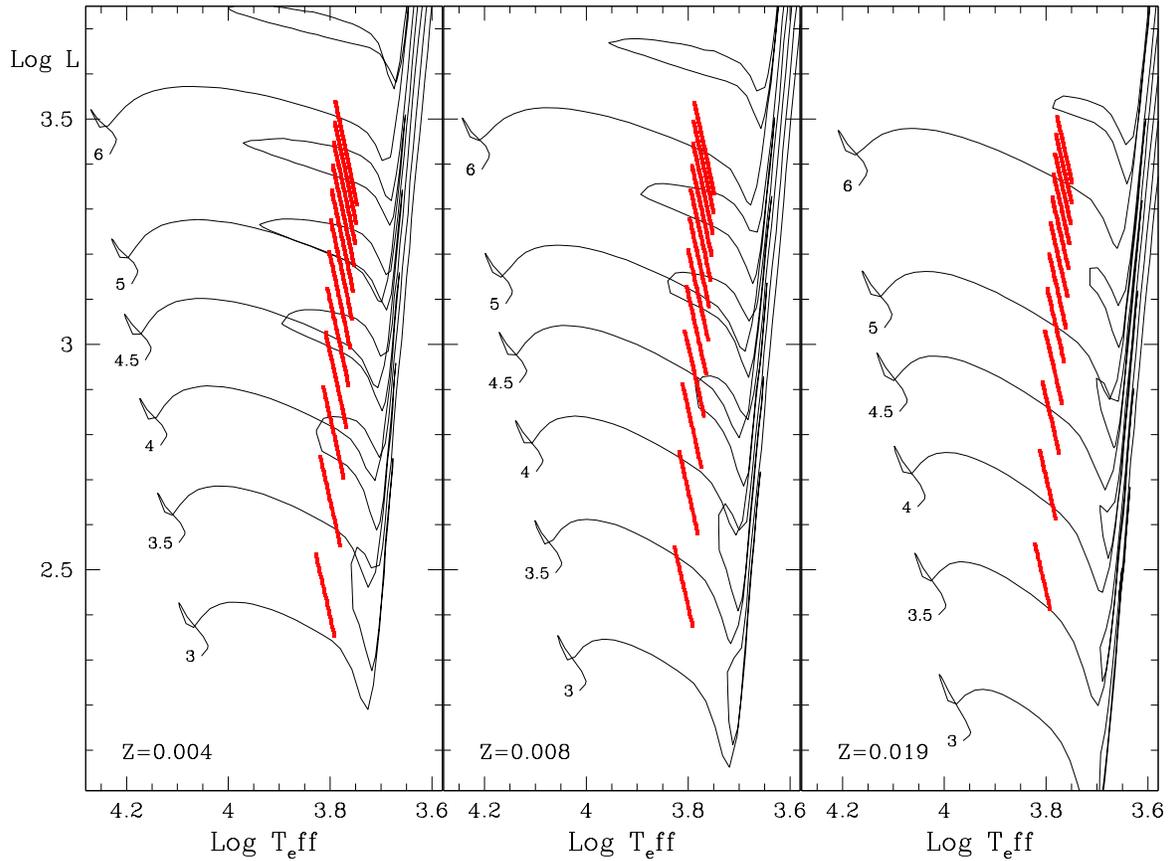}}
{\plotone{plotevol_GN93_bw.ps}}
\caption{\small Location of the Cepheid 
models with simultaneously 
linearly unstable F and O1 modes; the slanted lines are at constant
period, with $P_0$ ranging from 1.0 to 6.5\th d in steps of 0.5\th d; 
superposed are the Z=0.004, 0.008 and 0.019
 evolutionary tracks of \cite{girardi} for masses ranging from 3 to 6\th \Mo .
}
\label{figevolall}
\end{figure*}


\section{Linear Pulsation Models for Beat Cepheids}

Only the envelope of the Cepheids is engaged in pulsation.  This allows
theorists a convenient simplification: One needs to construct Cepheid envelopes
only on which one imposes an inner boundary condition (v=0) at a fixed core
radius $R=R_c$, where in addition the luminosity is constant $L= L_c$ and is
provided by the nuclear burning in the stellar interior, a region which is
irrelevant insofar as the pulsation is concerned.  The envelope which can be
taken as chemically uniform because of a prior evolutionary stage of extended
convection, is uniquely specified by 3 astrophysical parameters, \eg the
stellar mass $M$, the luminosity $L$ and the effective temperature \Teff\ of
the equilibrium model (\ie the state it would be in if it did not pulsate), as
well as the composition parameters, \ie $X$, $Y$ and $Z$, that appear in the
equation of state and in the opacities.  We recall that $X$ and $Y$ are the
mass fractions of hydrogen and helium, and that all other elements are lumped
into the 'metallicity' $Z$.  The three parameters, $M$, $L$, \Teff, are the
most convenient from a computational point of view, and are the input to our
equilibrium model builder and linear pulsation codes (\eg \cite{ykb98}).  Other
astrophysical parameters can be substituted, but at the expense of an
iteration, \eg the stellar radius of the equilibrium model, $R_*(L,M,\teff)$,
or one or more of the linear periods, $P_k(L,M,\teff)$, where $k=0,1,\ldots$
label the radial pulsation modes.  Note that the three basic parameters are
also the natural quantities $L(t)$, $M(t)$, and \Teff($t$) whose temporal
behavior, \ie the evolutionary tracks, are given by stellar evolution
calculations.

Our stellar equilibrium models are constructed and their linear (nonadiabatic)
periods $P_k$ and growth rates $\eta_k$ are computed with the Florida pulsation
code.  Turbulent convection is approximated with time-dependent mixing length
(described \eg in \cite{kbsc02,kbby98})\th \eg with the values of the $\alpha$
parameters chosen to be \{$\alpha_d$= 2.177, $\alpha_c$= 0.4, $\alpha_s$=
0.433, $\alpha_n$= 0.12, $\alpha_t$= 0.001, $\alpha_r$= 0.4, $\alpha_p$= 0,
$\alpha_\lambda$= 1.5\}).  OPAL \citep{iglesias96} and \cite{af94}
opacities are used.  

For our purposes we can use linear (nonadiabatic) periods.  Linear periods are
sufficiently accurate because they differ from the nonlinear ones at most in
the fourth decimal figure (\eg \cite{aa98}, \cite{sb07}).  It is in the modal
selection problem where nonlinearity plays an important role in limiting the
region where beat pulsations can occur (see \cite{kb01}, \cite{kbsc02}).  We
will address this issue in  \cite{sb07} which deals with nonlinear BC
model pulsations.

\begin{figure}
\epsscale{1.15}
\ifthenelse{\boolean{color}}
{\plotone{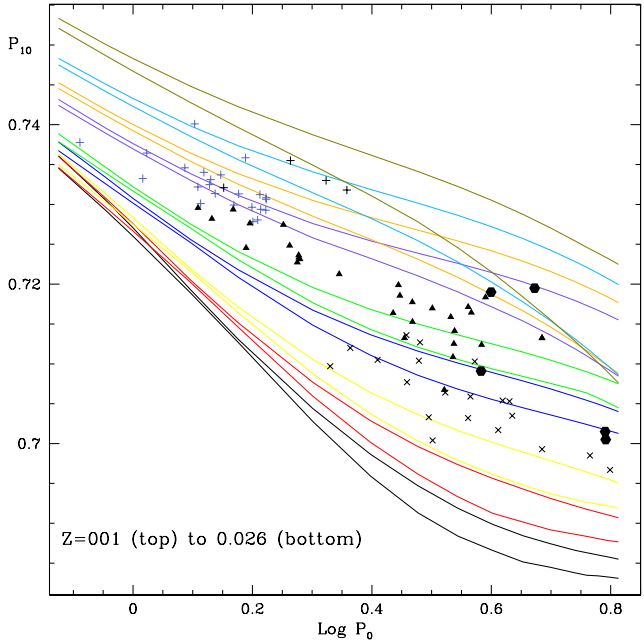}}
{\plotone{plotprat_edges_fGN93_bw.ps}}
\caption{\small $P_1/P_0$  \vs Log $P_0$  plot.  The pairs of lines delimit the
ranges for which both F and O1 are linearly unstable.  The metallicity
increases downward in the figure from $Z$ = 0.001, 0.002, 0.003, 0.004,
0.008, 0.010, 0.016, 0.020, 0.026.
For reference we have superposed the location of known Beat Cepheids (cf text).
}
\label{fig_prat_edges}
\end{figure}

\begin{figure*}
\epsscale{1.11}
\ifthenelse{\boolean{color}}
{\plotone{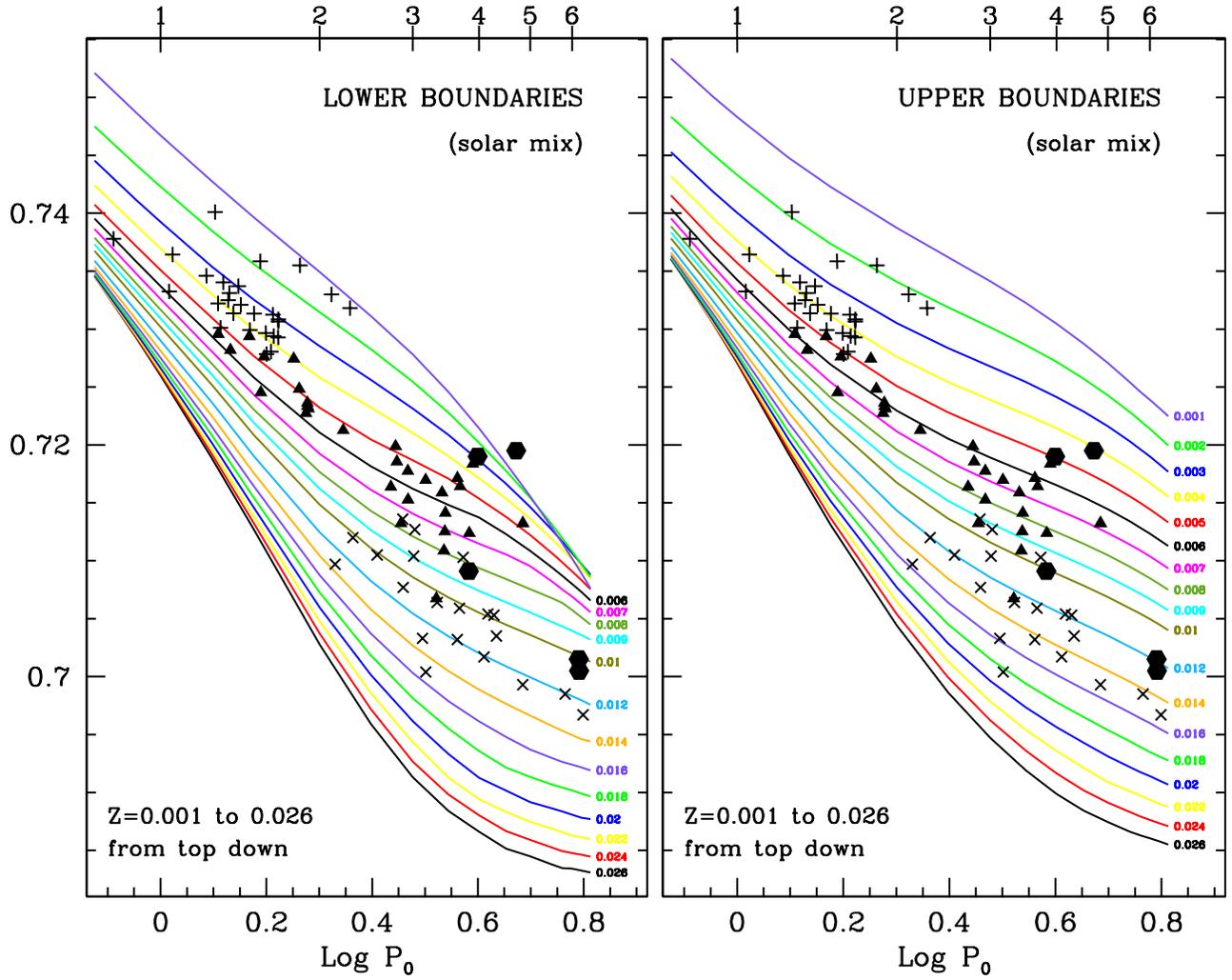}}
{\plotone{plotprat_edges_s_fGN93lab_bw.ps}}
\caption{\small Log $P_0$  \vs $P_1/P_0$ \th ($P_0$ on top axis).  
The lines delimit the
ranges for which both F and O1 are linearly unstable: Upper limit on the left
and lower limit on the right.  The metallicity
increases   downward in the figure  from $Z$ = 0.001 (top line) 
to 0.010 in steps 
of 0.001, and from 0.010 to 0.026 (bottom line) in steps of 0.002.
(A color figure with $\Delta Z$ =0.001 spacing everywhere 
is available with the electronic version.)
For reference we have superposed the location of known Beat Cepheids.
}
\label{fig_prat_edges_2}
\end{figure*}

During the galactic evolution metal enrichment occurs through helium and
hydrogen burning.  Because the helium content $Y$ thus correlates with $Z$, we
have assumed a relation $X=X(Z)$, obtained by fitting a parabola in Log Y \vs
Log Z through the 3 points ($Y$, $Z$)= (0.280, 0.020), (0.276, 0.008) and
(0.270, 0.004).  With this relation we reduce the 3 composition parameters
($X(Z)$, $Y(Z)$=1-$X(Z)$-$Z$, $Z$) to a single one, $Z$.


\section{Results}


\subsection{Log $M$ -- Log $L$ plane}

In their calculation of the metallicities of the BCs in the M33 galaxy
\citep{bbm06} used the 2 observed periods $P_0(L,M,\teff,X,Z)$ and
$P_1(L,M,\teff,X,Z)$, and the \Teff\ as the 3 parameters that determine the
stellar model.  This then produced, for a given composition, a one parameter
family of models, with \Teff\ as the parameter.  They then further limited the
models to those that were {\it linearly unstable simultaneously in the
fundamental {\bf F} and first overtone {\bf O1} modes}.  This yielded, for each
$Z$, loci of potential BCs in a Log $L$ \vs Log $M$ diagram \th (potential
because this double instability is only a necessary condition for BCs to
exist).

In Fig.~\ref{plotloci}, to illustrate the procedure, we similarly display the
$Z$ dependent set of loci of 6 hypothetical BCs, with the $(P_0, P_{10})$
values that are indicted in the figure.  For each sequence (in \Teff, with
fixed $P_0$ and $P_{10}$) we have plotted the loci as a function of $Z$.  For
the top 2 sequences on the left, $Z$ ranges from 0.003 to 0.017, and for the
other four it ranges from 0.001 to 0.020.

\begin{figure*}
\epsscale{1.15}
\ifthenelse{\boolean{color}}
{\plotone{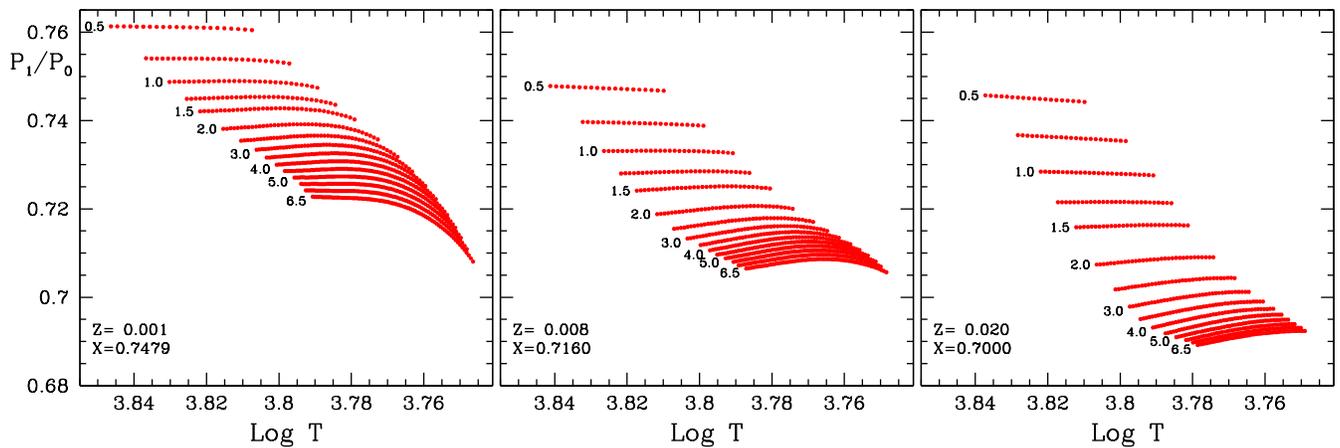}}
{\plotone{plot_t_pr_bw.ps}}
\caption{\small Period ratio \vs Log \Teff\ for models with simultaneously 
linearly unstable F and O1 modes, for 3 different compositions.  
The lines correspond to models with constant periods, running
from 0.5 to 6.5 days.}
\label{fig_t_pr}
\end{figure*}

\begin{figure}
\epsscale{1.18}
\ifthenelse{\boolean{color}}
{\plotone{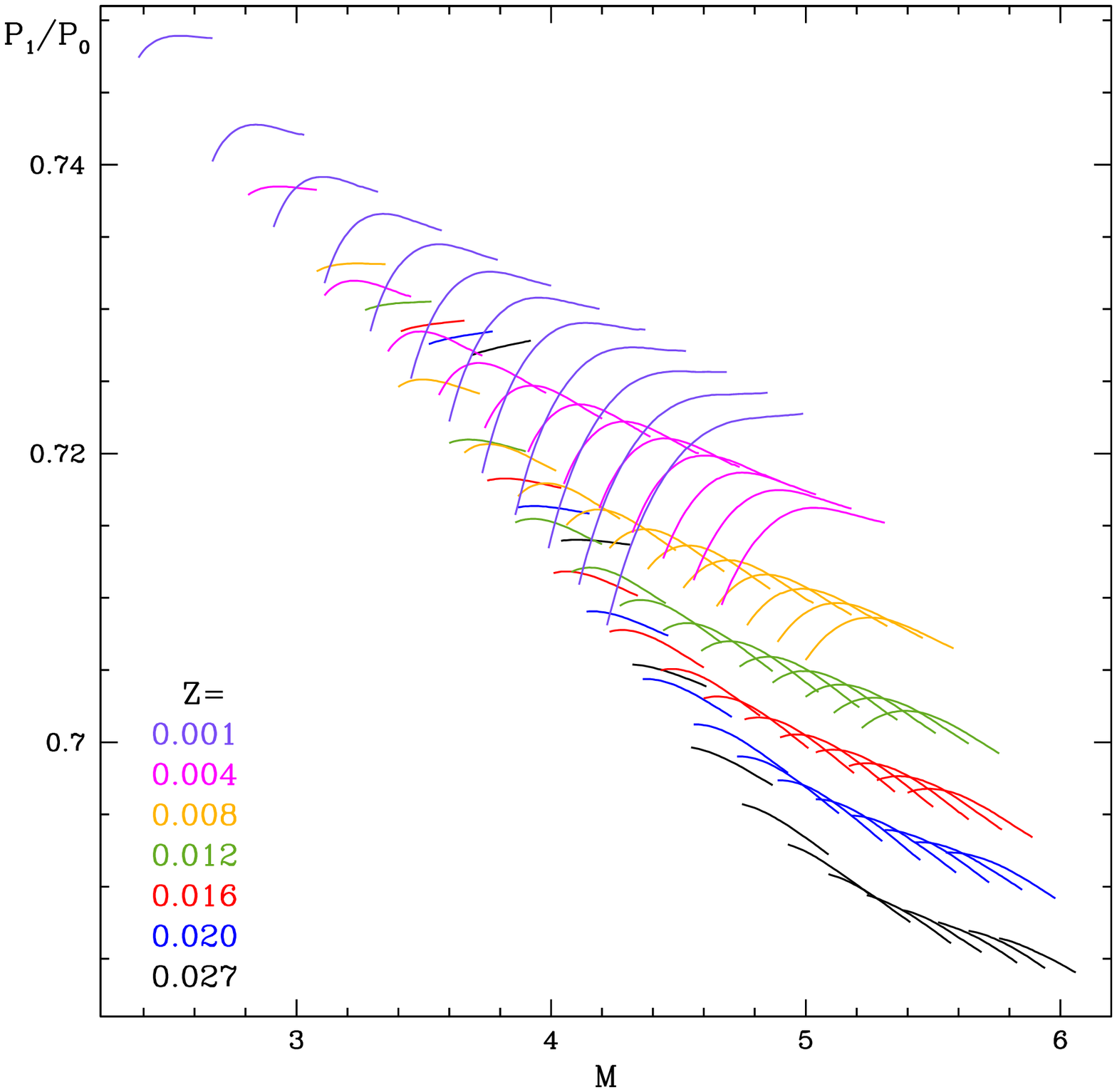}}
{\plotone{plot_m_pr_bw.ps}}
\caption{\small Period ratio $P_{10}$ \vs mass at constant period for BCs
candidates with the various values of $Z$ from the top down 
as indicated in the figure.  The
periods range from 1.0 to 6.5 days, with the larger masses going with the
larger periods.}
\label{fig_mpr}
\end{figure}

Superposed on the loci we display the \ML relations derived from \cite{piet},
for $Z$= 0.001, 0.004, 0.008, 0.019 and 0.027, from left to right.  The
acceptable BC candidate models are those for which the locus of a given $Z$
intersects the \ML curve for the {\it same} $Z$.  As noted in \cite{bbm06}, as
$Z$ increases the loci move toward the northeast, at first very slowly, then
they speed up as they cross the region of the \ML curves to slow down again
when they arrive on the right.  It is this rapid crossing of the \ML region
that allows one to narrow down the $Z$ of the BCs.  We recall
that the loci depend only on the period $P_0$ and period ratio $P_{10}$, and
pulsation theory.  Stellar evolution comes into play through the \ML relation
which determines what are the acceptable metallicity values for the given $P_0$
and $P_{10}$.  It is apparent from Fig.~1 of \cite{bbm06} that these values are
relatively insensitive to the chosen \ML relation (see also \S\th\ref{sec:ml}
below).

While this procedure allows us to determine the metallicity of BCs, it is
computer intensive as it necessitates the computation of the loci for a range
of $Z$, and this for each BC.  Furthermore for some parameter values that fall
far away from those of normal Cepheids the models do not always converge.


\subsection{Log $P_0$ -- $P_{10}$  plane (Petersen Diagrams)}

We find it useful to present a different, more practical procedure.  This
consists of constructing Petersen diagrams ($P_{10}, P_0\th ; Z)$ in which we
delimit the ranges of acceptable BC models for a finely spaced set of $Z$
values.  For a given BC with its observationally derived $P_0$ and $P_{10}$ one
can then readily determine into which of these ranges it falls, interpolate,
and thus impose upper and lower limits on $Z$.

Our PDs are constructed as follows.  We specify the F period
$P_0(L,M,\teff,Z)$, and the \ML relation $L=L(M,Z)$.  These 2 constraints allow
us to compute a sequence of models of a specified $Z$ with different values of
\Teff\, and consequently different $P_{10}(M,L,\teff,Z)$.  Of these models we
then only retain those that are linearly unstable both in the F and the O1
mode.  Our set of $P_0$ ranges from 0.75\th d to 6.5\th d.  Our $Z$ values
range from 0.001 to 0.026 in steps of $\Delta Z$ = 0.001.

We recall that there exists a slightly slanted, vertical strip in the HR
diagram, called the instability strip ({\bf IS}) in which the stars become
unstable to pulsation and are called Cepheids.  In Figure~\ref{figevolall} we
display our doubly unstable models in an HR diagram for selected metallicities
($Z$=0.004, 0.008 and 0.019).  The lines run at an angle because the sequences
(in \Teff) are computed at constant period $P_0$ and constant $P_{10}$.  The
full IS is much wider in \Teff\ as its extends leftward and rightward into the
regimes where {\sl only} the O1 and F modes, respectively, are linearly
unstable.  Note also that not the whole doubly unstable region will give rise
to BCs:~~first, simultaneous instability in the two modes is only a necessary
condition for beat behavior, and second, the tracks that actual Cepheids follow
may indeed not penetrate everywhere into this region.  In these calculations we
have made use of a \ML relation derived from \cite{girardi} shown in Table.~1
(we will return to a discussion of \ML relations in \S\ref{sec:ml}).

In Figure~\ref{figevolall} we superpose the tracks of \cite{girardi} for
Z=0.004, 0.008 and 0.019, respectively.  Among the many tracks in the
literature we have chosen these because they cover the whole mass range and $Z$
range of interest to us.  Furthermore we have found that the pulsational
properties of the Cepheid models along these evolutionary tracks, as calculated
with our pulsational code, are in good agreement with the resonance constraints
imposed by the OGLE-2 data of the Small and Large Magellanic Clouds
\citep{bkb04}.  

\begin{figure}
\epsscale{1.15}
\ifthenelse{\boolean{color}}
{\plotone{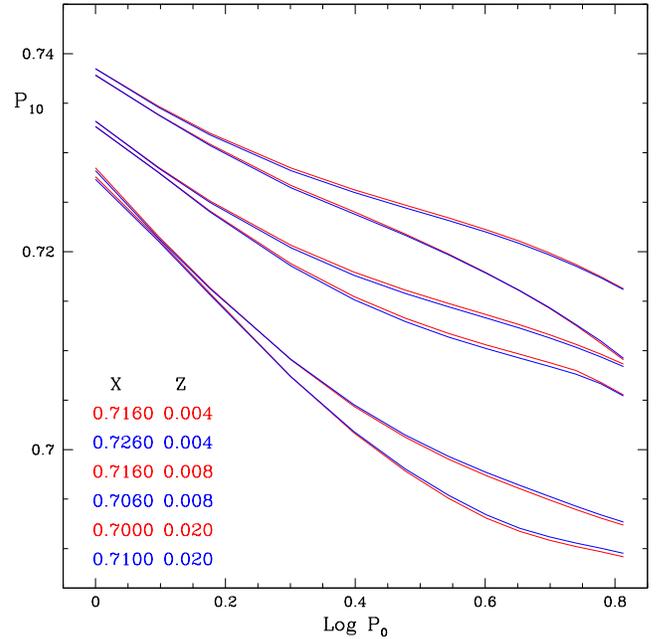}}
{\plotone{plotprat_edges_y_bw.ps}}
\caption{\small $P_1/P_0$ \vs Log $P_0$ plot.   Effect of changing the helium
content $Y$ (=1--$X$--$Z$) at fixed $Z$. The pairs of lines delimit the
ranges for which both F and O1 are linearly unstable.
}
\label{fig_prat_edges_y}
\end{figure}

The key results of our survey are displayed as a PD in
Fig.~\ref{fig_prat_edges}.  The pairs of lines delineate for each $Z$ the range
of $P_{10}$ values for which the F and O1 modes are simultaneously linearly
unstable.  The metallicity is increasing downward for selected values of $Z$ =
0.001, 0.002, 0.003, 0.004, 0.008, 0.010, 0.016, 0.020, 0.026.  The lines
become crowded for the higher $Z$ indicative a spacing that is closer to Log
$Z$ than to $Z$.

Because of overcrowdedness we could only display very few values of $Z$ in
Fig.~\ref{fig_prat_edges} which we have therefore split into 2 subfigures 
Fig.~\ref{fig_prat_edges_2}, showing respectively the lower and upper limits of
$P_{10}$ for a finer set of $Z$ values.  The $Z$ are increasing downward from
$Z$=0.001 to 0.010 insteps of 0.001, and then from $Z$ = 0.010 to 0.026 in
steps of 0.002.

For reference we have superposed the location of known BCs in the Galaxy (x's),
in M33 \citep{bbm06} (hexagons), in the LMC (triangles)
\citep{alcock95,soszynski00} and in the SMC (crosses)
\citep{beaulieu97,udalski99}.

Fig.~\ref{fig_prat_edges} and, more conveniently, Fig.~\ref{fig_prat_edges_2}
can now be used to infer upper and lower bounds on the metallicity of a given
BC star from its coordinates, (${\rm Log}P_0$, $P_{10}$).  We identify the
range(s) into which a star falls which gives a $Z_{min}$ and $Z_{max}$.  If it
falls between the calculated ranges, as happens for short periods, we get a
$Z_{min}$ and $Z_{max}$ from the adjacent ranges.  One can refine $Z_{min}$ and
$Z_{max}$ by interpolation in the table from which the figures are constructed.

Consider, for example, the M33 BC (hexagon) with the lowest $P_{10}$.
The left panel of Fig.~\ref{fig_prat_edges_2} shows that the star falls
between the $Z$=0.010 and 0.011 lines, and from the right we find
0.012$<Z<$0.013.  An interpolation gives  0.0106 $<Z<$ 0.0124 with an average
of $Z$=0.0115.

The 'Beat Cepheid metallicities' that we infer for the Galactic BCs, $Z$=0.0074
-- 0.0182, with an average of $Z$=0.0118 are on the low side, but for the MC and
the M33 BCs they fall in the generally accepted ballpark, 0.001 -- 0.007 with
average $Z$ = 0.0037 for the SMC, 0.0035 -- 0.012, with average = 0.0062 for
LMC, and 0.0075 -- 0.0124, with average = 0.0075 for M33.  We will return to a
discussion of a possible remedy for this discrepancy.

In the upper left corner of Fig.~\ref{fig_prat_edges} , the ranges for the
plotted $Z$ values do not overlap, meaning that the position in the PD very
narrowly determines the allowed $Z$ for a given $P_0$ and $P_{10}$.  In the
upper right, in contrast, there is more overlap and the allowed range of $Z$ is
a little broader, albeit still in the $\pm 0.001$ range for $Z$.  In general
toward the bottom there is less overlap as well.

Fig.~\ref{fig_prat_edges_2} provides us with no information about how the
temperature or mass vary between the upper and lower loci at fixed $P_0$.  In
Fig.~\ref{fig_t_pr} we display the behavior of the period ratio $P_{10}$ as a
function of \Teff, each sequence at fixed period $P_0$.  We see that for high
$Z$ (on the right) the upper boundary, \ie the maximum $P_{10}$, of the doubly
unstable models is at a higher \Teff\ for the short periods, but the reverse is
true for the longer periods.  For the low $Z$ (left) the behavior of $P_{10}$
with \Teff\ is not even monotone.  Referring back to Fig.~\ref{fig_prat_edges}
it is therefore apparent that the behavior of \Teff\ between the boundaries of
the doubly unstable models (vertically, at fixed $P_0$) can be increasing,
decreasing or even be nonmonotone.  Fig.~\ref{fig_mpr} indicates that similar
nonmonotone behavior occurs for $P_{10}$ as a function of mass.

In the following we examine what physical factors can influence the PD, and we
will examine in turn the effects of the helium mass fraction $Y$ of the \ML
relation, and of stellar rotation.


\subsection{Effect of the Helium Content $Y$}

Because we have chosen the helium content relation $Y(Z)$ in a somewhat
cavalier way, we want to explore the sensitivity of the ranges of BC models in
the PD to $Y$.  In Fig.~\ref{fig_prat_edges_y} we show the ranges for 3
selected values of $Z$ = 0.004, 0.008 and 0.20.  The figure also displays the
associated $X$ and $Y$ values.  It is very clear from this figure that {\it the
sensitivity to even large changes of $Y$ is minimal}.  The period ratios are
essentially only sensitive to $Z$ (because opacity mostly is mostly sensitive
to $Z$).


\begin{figure}
\epsscale{1.15}
\ifthenelse{\boolean{color}}
{\plotone{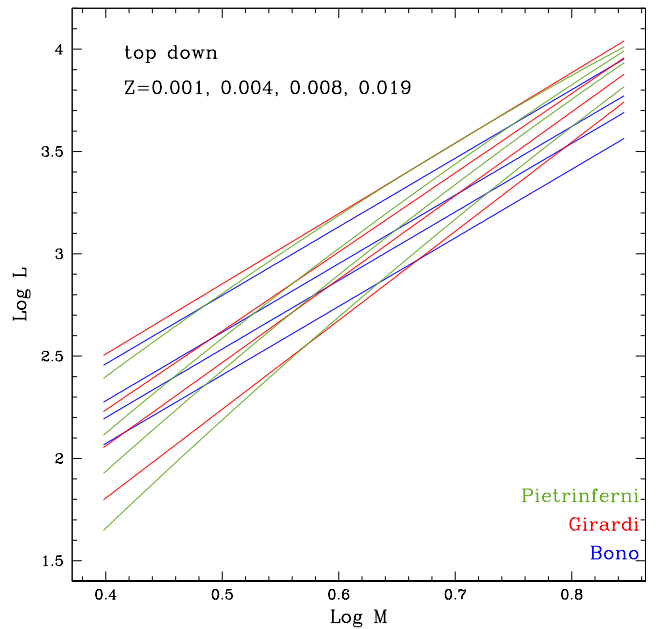}}
{\plotone{plot_ml_bw.ps}}
\caption{\small \ML relations:  $Z$ =0.001, 0.004,
0.008, 0.019, from the top down.
}
\label{plotml}
\end{figure}

\begin{figure}
\epsscale{1.15}
\ifthenelse{\boolean{color}}
{\plotone{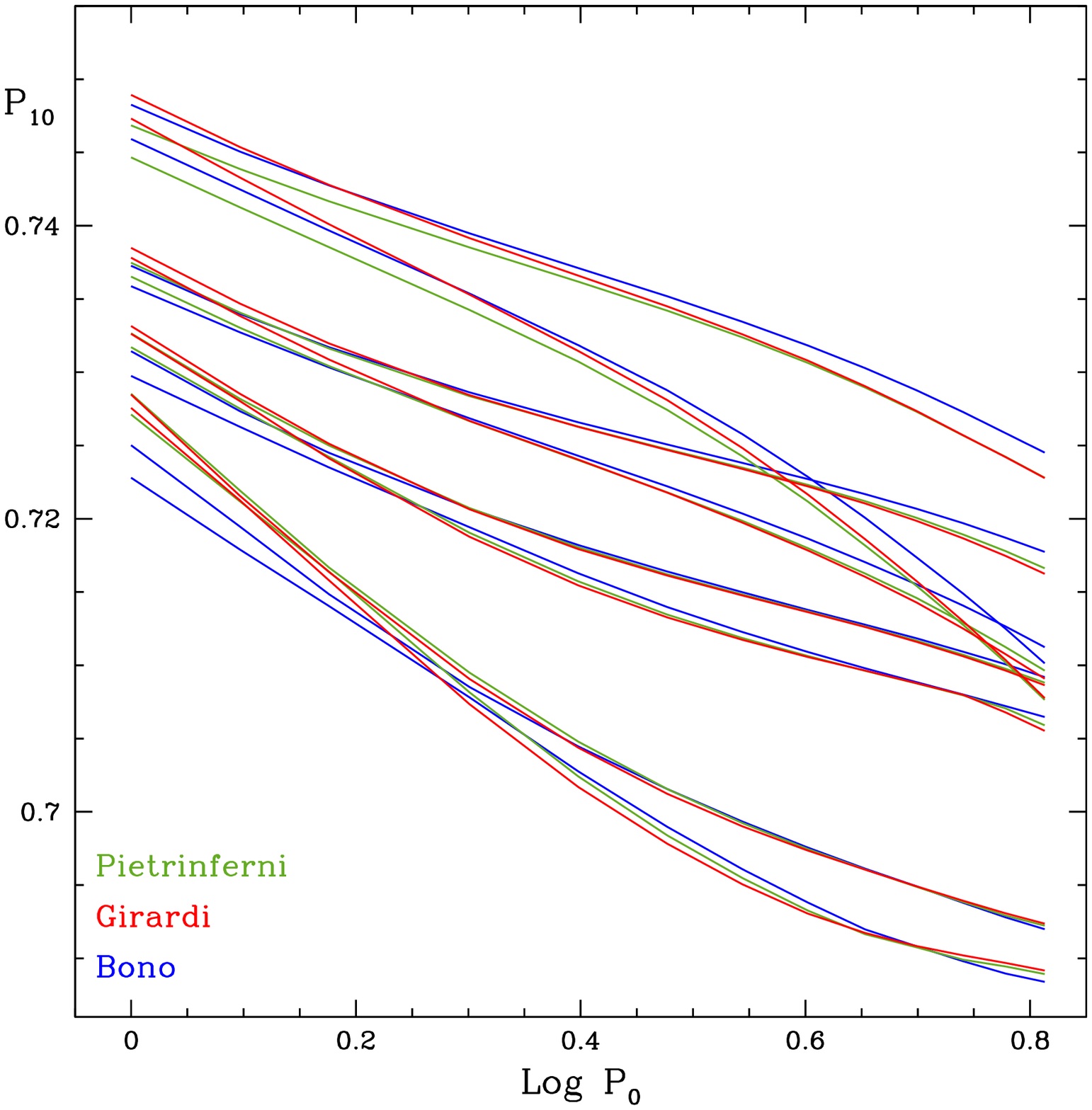}}
{\plotone{plotprat_edges_girbonpie_bw.ps}}
\ifthenelse{\boolean{color}}
{
\caption{\small $P_1/P_0$ \vs Log $P_0$ plot.  Effect of the \ML relation:
\cite{girardi} (black),  \cite{piet} (red) and  Bono \etal (blue).
The pairs of lines delimit the
ranges for which  F and O1 are simultaneously linearly unstable.}
}{
\caption{\small $P_1/P_0$ \vs Log $P_0$ plot.  Effect of the \ML relation:
\cite{girardi} (solid),  \cite{piet} (dashed) and  Bono \etal (dotted).
The pairs of lines delimit the
ranges for which  F and O1 are simultaneously linearly unstable. 
}
}
\label{fig_prat_edges_girbonpie}
\end{figure}

\subsection{Effects of the \ML relation}  \label{sec:ml}

It is well known that a Cepheid star of given mass may cross the IS three
times, as seen \eg in Fig.~\ref{figevolall} for the \cite{girardi} tracks.  The
first crossing occurs very fast, and, statistically, very few stars would be
caught in that state of evolution.  Most Cepheids are expected to be found on
the slow leftward (blueward) loop, and some on the faster rightward return.

Clearly the tracks of Fig.~\ref{figevolall} show some serious and well known
deficiencies as far as observation is concerned (which are shared by all other
published evolutionary tracks):~~the blueward 'Cepheid' loops peter out as $Z$
is increased, and by the time one reaches Galactic values of $Z$, there are no
more loops at all that penetrate the BC Cepheid region, in strong discordance
with observation.  While some decrease of the lengths of the horizontal loops
with $M$ is acceptable because there is a lower observed cutoff in the Cepheid
periods, it is patently excessive.  A good discussion of the problems
associated with the horizontal Cepheid loops can be found in
\cite{cordierthesis} and \cite{cordier02}.  
 
Aside from these problems, the very use of an \ML relation is an approximation,
first, because we limit ourselves to the second crossing, and second, because
even then the stellar tracks are not horizontal in the HR diagram ($L$-\Teff\
plane).  In addition, referring back to Fig.~\ref{figevolall} one realizes that
extracting an \ML relation from the evolutionary tracks requires some guesswork
as to where some of the aborted horizontal loops really should be.  Another
problem is that in essentially all the evolutionary calculations the horizontal
loops exhibit a nonmonotone behavior with $Z$, some of which may be of
numerical origin.
There is also broad evidence from the LMC and SMC Cepheids that the computed
evolutionary tracks may be underluminous for a given mass to account for their
pulsational properties (\eg \cite{bbk01}), and a smilar conclusion was reached
for the bump Cepheids by \citep{kw02,bono02}.  However, it is not our purpose
here to present a critical review of the various evolutionary calculations and
the \ML relations one can derive from them, especially, as it will turn out
that our application to BCs fortunately has only a moderate sensitivity to the
\ML relation.  This was already noted in \cite{bbm06} and in
Fig.~\ref{plotloci} above.

With these caveats, using the blue tips of the \cite{girardi} tracks we have
made fits to Log $L$ that are bi-quadratic in Log M and Log $Z$ in the range
$2.5 < M < 6.5$ and $0.004< Z <0.02$.  They have the largest uncertainty for large
$Z$ in general, and for small $M$ (because of the lack of horizontal loops in
the evolutionary tracks).  Table~1 presents the \ML relations that we have
extracted from \cite{girardi} and from \cite{piet}.  Both these evolutionary
calculations give a full grid of M values and Z values in the range that we are
interested in, and they use a nonzero overshooting parameter.  \cite{alibert}
give an \ML only for some values of $Z$ and we have therefore not used them
here.  \cite{cordier02} find that the evolutionary calculations without
overshooting does not give good agreement with observational constraints.  In
addition, we find that for Cepheids with periods as low as 15\th d with Bono's
\ML relation one would need to go to very high masses of 8.5 -- 9\th \Mo\ as
opposed to 7.0 -- 7.5\th\Mo\ for the other \ML relations.  However, just for
comparison we also consider the \ML of \cite{bono00} derived from tracks without
any overshooting.  \ML fitting formulae are presented in their paper.

Fig.~\ref{plotml} displays the \ML relations of \cite{girardi,piet,bono00} for
$Z$ = 0.001, 0.004, 0.008 and 0.019.  One sees that Bono is much flatter than
the other two.

The period ratios $P_{10}$ obtained with the three \ML relations are compared
Fig.~\ref{fig_prat_edges_girbonpie}.  The metallicities range from $Z$=0.001
(top), 0.004, 0.008 to 0.020 (bottom).  Despite the large difference in the 3
\ML relations, one finds, perhaps unexpectedly, that the PDs for the 3
relations agree quite well for periods down to $P_0$ = 1.5\th d for all $Z$.
The discrepancies between the \ML relations show up for short periods only, and
especially for Bono.

\begin{table}
\caption{\small \ML relations}
\vspace{-5mm}
\begin{center}
$Log L = b_1 + b_2 Log M + b_3 Log Z +b_4 Log^2 M + b_5 Log^2 Z
+b_6 Log Z Log M$
\medskip
\begin{tabular}{r r r r r r r}
\hline\hline
    \noalign{\smallskip}
& $b_1$ & $b_2$& $b_3$ & $b_4$& $b_5$ & $b_6$ \\
    \noalign{\smallskip}
\hline
    \noalign{\smallskip}
Gir& 
--3.2822 &
  8.0864  &
--1.7988 &
--1.2764 &
--0.1771 &
  0.9588 \\
    \noalign{\smallskip}
\hline
    \noalign{\smallskip}
Piet&
--2.0937 &
  5.5361 &
--1.5061 &
  0.0306 &
--0.1421 &
  0.7112 \\
    \noalign{\smallskip}
\hline
\end{tabular}
\label{tab:beat}
\end{center}
\end{table}

\begin{figure}
\epsscale{1.15}
\ifthenelse{\boolean{color}}
{\plotone{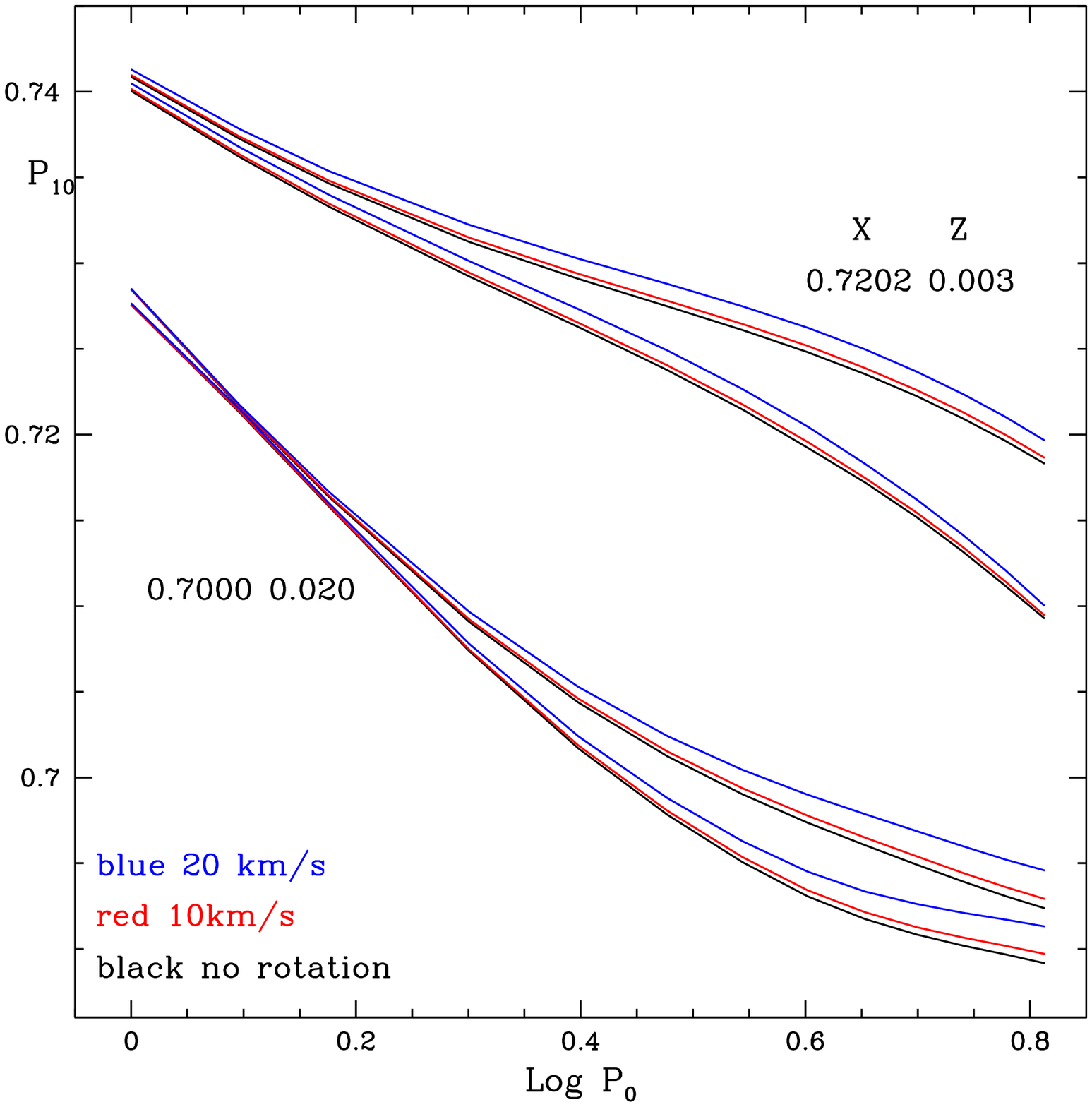}}
{\plotone{plotprat_edges_rot_bw.ps}}
\caption{\small $P_1/P_0$ \vs Log $P_0$ plot.  Effect of the rotation.  
 The pairs of lines delimit the
ranges for which both F and O1 are linearly unstable. 
The
curves are for $v_{rot}$ = 0, 10\th and 20\th km/s.  
Note that the effect of rotation is most likely overestimated 
by the treatment as a pseudo-rotation.
}
\label{fig_prat_edges_rot}
\end{figure}

\begin{figure}
\epsscale{1.15}
\ifthenelse{\boolean{color}}
{\plotone{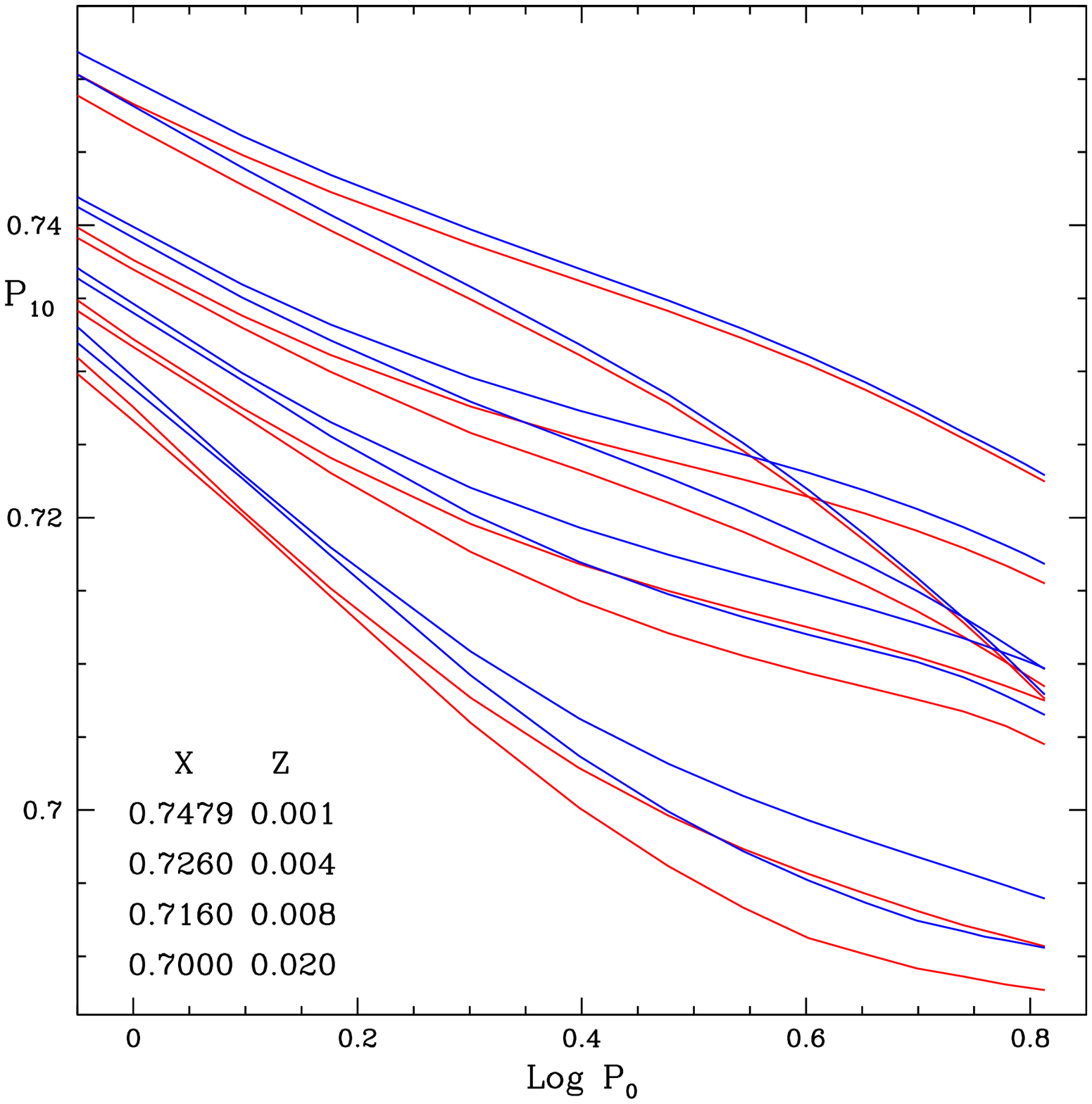}}
{\plotone{plotprat_edges_GN93MFE75_bw.ps}}
\ifthenelse{\boolean{color}}
{\caption{\small $P_1/P_0$ \vs Log $P_0$ plot.  Effect of changing the chemical
makeup of $Z$: heavier elements  reduced by a factor of 0.75 (blue), which
shifts the curves upward to higher $P_{10}$.}  
}
{\caption{\small $P_1/P_0$ \vs Log $P_0$ plot.  Effect of changing the chemical
makeup of $Z$: heavier elements  reduced by a factor of 0.75 (dashed), which
shifts the curves upward to higher $P_{10}$.}  
}
\label{fig_prat_edges_GN93M75}
\end{figure}

\begin{figure*}
\epsscale{1.11}
\ifthenelse{\boolean{color}}
{\plotone{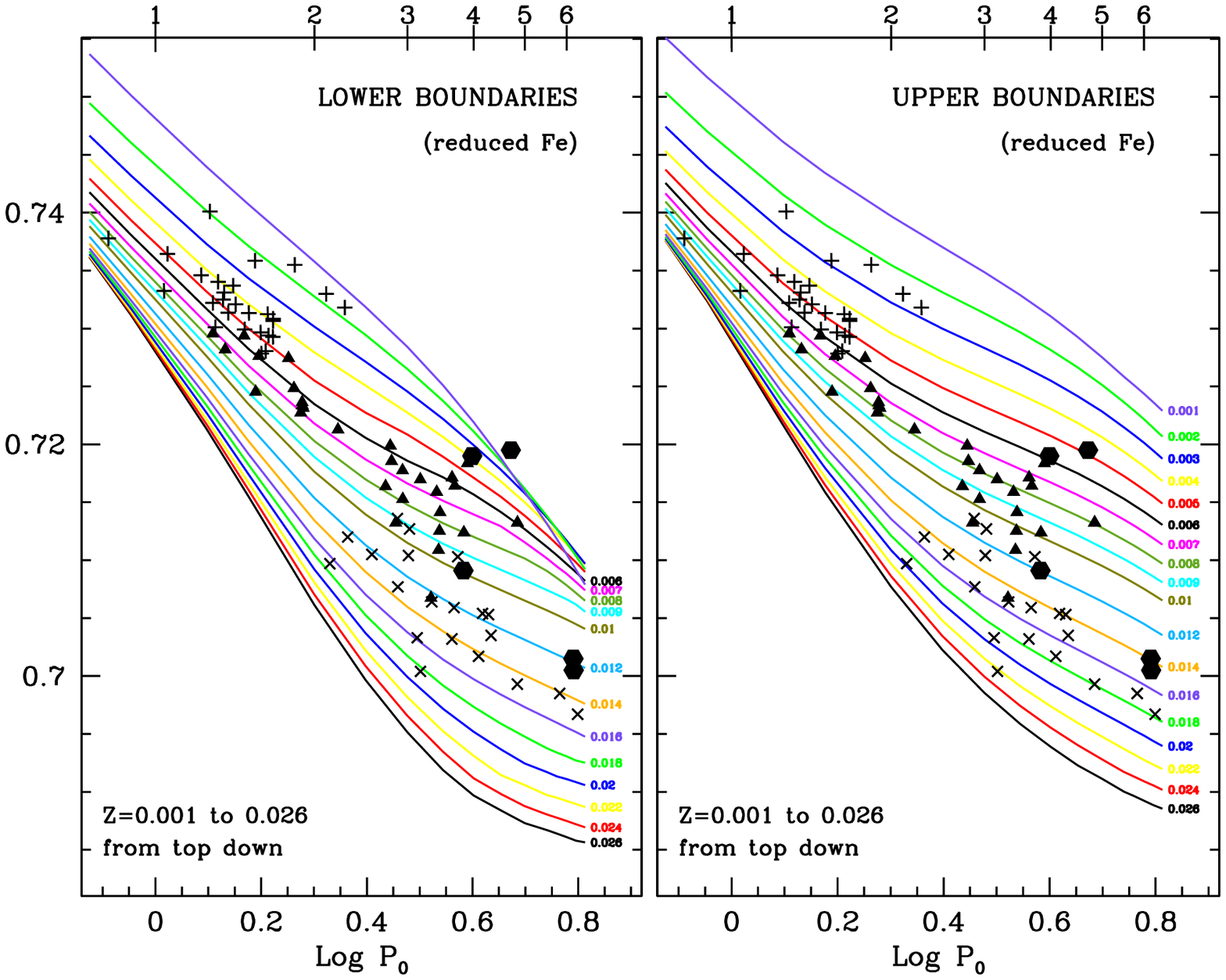}}
{\plotone{plotprat_edges_s_MFE75lab_bw.ps}}
\caption{\small $P_1/P_0$ \vs Log $P_0$ plot \th ($P_0$ on top axis).  
Effect of changing the chemical
makeup of $Z$: heavier elements reduced by a factor of 0.75\th.}  
\label{fig_prat_edges_specM75}

\vskip 5mm

\end{figure*}


\subsection{Effects of Rotation}

The effect of rotation on PDs has been considered in the past, generally with
the help of the crude, but reasonable simplifying assumption of a 'spherical
centrifugal force', $F = \omega^2 r$ (\cite{stothers}; \cite{kb90}) that
is added in the model builder and in the linear stability analysis.  Note that
this subterfuge is likely to exaggerate the effect of rotation.

The results of our calculations with pseudo-rotation are displayed in
Fig.~\ref{fig_prat_edges_rot} for surface rotation velocities of $v_{rot}$ =
0, 10\th km/s and 20\th km/s.  The PD indicates only small shifts even
with rotations as high as 20\th km/s.  A comparison with
Fig.~\ref{fig_prat_edges} shows that for the low $Z$ models, with their high
$P_{10}$ values, the neglect of rotation amounts up to 20\th km/s generally
leads to an error in $Z$ of less than 0.001, but up to 0.002 for the longer
periods and smaller $P_{10}$ (larger $Z$).

Recently a flag has been raised by \cite{sgg06} who found that the PDs of delta
Scuti models exhibit a large sensitivity to the assumed rotation rate, and that
even a modest rotation rate can cause an increase in the period ratio which
would then be misinterpreted as a lower metallicity than the star has.  In
contrast, we have seen that for Cepheid models rotation has only a small effect
on PDs, all the more so that the observed Cepheid rotation rates ($v_{rot}$ sin
$i$) are typically less than 20\th km/s \citep{nardetto} and thus much smaller
than those of delta Scuti stars.

We conclude that neither the helium mass fraction $Y$, nor the stellar
rotation, nor the \ML relation can appreciably affect the PD.  Remaining
uncertainties are associated with the relative elemental concentrations of the
'metals' in $Z$.

\subsection{Effect of Elemental Mixture that is Lumped into $Z$}

So far we have used the OPAL \citep{iglesias96} opacities with the standard
solar elemental mixture of \cite{gn93}.  One needs to ask the question of
whether the PDs are sensitive to $Z$ as a whole, or to the relative fraction of
the heavier metals, \ie the Fe group elements, which provide the bulk of the
opacity.  We therefore have made several tests.

In Figs.~\ref{fig_prat_edges_GN93M75} and \ref{fig_prat_edges_specM75} we
display PDs obtained with OPAL opacities in which the number densities of the
dominant heavy elements (Fe, Ca, Ar, S, Si and Mg) are arbitrarily lowered by
factors of 0.75 \th (before renormalization to unity) in a 'Type 1 OPAL Table'
with fixed $X$, $Y$ and $Z$.  This can be compared to the PD of
Fig.~\ref{fig_prat_edges} which was for a solar mix (GN93 in OPAL).  One notes
that, in first approximation, the lines are just shifted, but there is also a
deformation.  For example, the old $Z$=0.022 line in Fig.~\ref{fig_prat_edges}
is now the $Z$=0.026 line in Fig.~\ref{fig_prat_edges_specM75}, and the
$Z$=0.005 is close to the 0.006 line.
 
As far as the Galactic BCs are concerned we find a range of $Z$ = 0.0089 --
0.0221, with an average of $Z$= 0.0141, better than the average of
0.0118 that we find with a solar mix, but still a little low compared to the
generally quoted values.  For the SMC we have $Z$ = 0.0015 -- 0.0100 with
average = 0.0047, for the LMC 0.0047 -- 0.0148, average = 0.0076, and for M33
0.0013 -- 0.0146, average = 0.009.

We are led to conclude that there is a discrepancy between the stellar models
that use opacities with the GN95 solar mix and the metallicities that are
quoted in the literature.  Better agreement can be achieved with a reduction in
the relative number densities of the Fe group elements.

We have also examined other changes.  The Grevesse G91 mix (in OPAL) has a
reduction of about 10\% in the heavier elements. Not astonishingly, the results
fall in between those of the 25\% reduction and GN95.  Similarly, increasing
the heavy elements (Fe, Ca, Ar, S, Si and Mg) by a factor of 1.1 produces a
comparable change, but in the opposite direction, as expected.  The
alpha-enhanced opacities, labeled Charboyer in the OPAL opacity library, which
have a chemical makeup that is quite different from solar produce large changes
in the PD.  Space does not permit us to look with more detail into the effects
of the chemical makeup in order to see if we can isolate what effects the
various subgroups of elements in the mix have on the PDs.  This will be done in
a subsequent paper.

We conclude that the chemical makeup does play a non-negligible role, mostly
for the short periods.  Decreasing the relative Fe content at fixed $Z$ is seen
to be equivalent to decreasing $Z$ at fixed elemental mixture in first
approximation.  This confirms what one may have expected, namely that the PD
are predominantly sensitive to the strength of the opacity which is provided by
the Fe group elements.


\section{Conclusions}

With the help of our convective pulsation code we have constructed Petersen
diagrams ($P_{10}$ \vs Log $P_0$) for Beat Cepheids.  In these PDs we have
delineated the region of simultaneous linear instability in the F and the O1
modes for a range of relevant values of metallicity $Z$.  Our results are
presented in Fig.~\ref{fig_prat_edges_2} such a way that from the measured
period $P_0$ and period ratio $P_{10}$ one can determine upper and lower limits
on the metallicity $Z$ of the corresponding Beat Cepheid.

The results are found to be rather {\it insensitive} (1) to the assumed helium
content $Y(Z)$, (2) to the exact form of the mass-luminosity relation, and (3)
to the presence of stellar rotation.  By comparing PDs with solar abundances to
those in which the heavier elements (in particular Fe) are reduced or increased
relative to the lighter elements we conclude that the PDs are
sensitive to the chemical makeup of $Z$.

Interestingly, we find that with a reduction in the number densities of heavier
elements at fixed $Z$ the 'pulsation metallicities' for the Galactic Beat
Cepheids are in better agreement with the values that are in the literature
than with the solar elemental mix of Grevesse-Noels (1993).  We plan to examine
the effect of the relative abundances in greater detail in a companion paper,
especially in light of recent suggested revisions of the abundances
\citep{ags05}.

\begin{acknowledgements}

We thank Daniel Cordier for giving us an update on the state of stellar
evolution calculations, Peter Wood, Jean-Philippe Beaulieu, Zolt\'an
Koll\'ath, L\'aszl\'o Szabados and Marie-Jo Goupil for valuable discussions,
and P. Wils for drawing our attention to the ASAS3 data on DZ CMa.  JRB
profited from the 2006 workshops on galaxies at the Aspen Center for Physics.
This work has been supported by NSF (OISE04-17772 and AST03-07281) at UF.  RSz
also acknowledges the support of a Hungarian E\"otv\"os Fellowship.

\end{acknowledgements}


\end{document}